\documentclass[aps,prd,showpacs,twocolumn,showkeys,preprintnumbers,floatfix,nofootinbib]{revtex4}

\usepackage{graphicx}                                 
\usepackage{amsmath,amsfonts}
\usepackage{bm}                                       
\graphicspath{{./Figures/}}                           

\newcommand{\op}{\langle\bar{\psi}i\gamma_{5}\tau^{3}\psi\rangle}  
\newcommand{\ReL}{\mathop{\mathrm{Re}L}}                            
\newcommand{\chiL}{\chi_{\textrm L}}                               

\newcommand{\etal}{\textit{\mbox{et al.\ }}}                       
\newcommand{\ie}{\textit{\mbox{i.e.\ }}}                           
\newcommand{\eg}{\textit{\mbox{e.g.\ }}}                           

\newcommand{\Fig}[1]{Fig.~\ref{#1}}
\newcommand{\Tab}[1]{Tab.~\ref{#1}}

\begin{document}


\preprint{HU--EP--05/78} \preprint{ZIB-Report 05-48} \preprint{SFB/CPP-05-81} 
\title{Probing the Aoki phase with \bm{$N_f=2$} Wilson fermions
  at finite temperature}

\author{E.--M.~Ilgenfritz, W.~Kerler, M.~M\"uller--Preussker,
  A.~Sternbeck} 
\affiliation{Humboldt-Universit\"at zu Berlin, Institut f\"ur Physik,
  D-12489 Berlin, Germany}

\author{H.~St\"uben}
\affiliation{Konrad-Zuse-Zentrum f\"ur Informationstechnik Berlin,
  D-14195 Berlin, Germany}

\date{November 28, 2005}

\begin{abstract}
   In this letter we report on a numerical investigation of the Aoki phase
   in the case of finite temperature which continues our former study at
   zero temperature. We have performed simulations with Wilson fermions
   at $\beta=4.6$ using lattices with temporal extension
   $N_{\tau}=4$. In contrast to the zero temperature case, the existence
   of an Aoki phase can be confirmed for a small range in $\kappa$ at
   $\beta=4.6$, however, shifted slightly to lower $\kappa$. Despite
   fine-tuning $\kappa$ we could not separate the thermal transition
   line from the Aoki phase.
\end{abstract}

\keywords{Wilson fermions, phase diagram, parity-flavor symmetry, Aoki
          phase, finite temperature}
\pacs{11.15.Ha, 12.38.Gc}
\maketitle

\section{Introduction} \label{sec:intro}

The interest in the phase structure of Wilson fermions coupled to
$SU(3)$ gauge fields has revived recently, in particular due to the
unexpected discovery of a first-order transition in twisted mass
QCD~\cite{Farchioni:2004us}. This transition survives when the Wilson
gauge action is replaced by a renormalization group improved gauge
action~\cite{Farchioni:2004fs}.  This has brought reports on a
nontrivial phase structure ~\cite{Blum:1994eh,Aoki:2004iq} back into
the center of interest including the pioneering paper by
Aoki~\cite{Aoki:1983qi}.

The \emph{Aoki phase}, characterized by a non-vanishing condensate $\op$ and
a broken pion mass triplet, was analyzed in the framework of chiral
perturbation theory in~\cite{Sharpe:1998xm,Munster:2004am}.
In~\cite{Munster:2004am} two scenarios were described: at a given
gauge coupling $\beta$ either the Aoki phase is realized in a certain
$\kappa$ interval (separated by second order phase transition lines
from a phase with degenerate massive pions), or there are first order
transitions in a certain interval of twisted masses (also ending at
second order transition points).

In a recent paper~\cite{Ilgenfritz:2003gw} we have found evidence
that the Aoki phase at zero temperature is unlikely to extend 
to \mbox{$\beta > 4.6$}, whereas the mentioned new phase transition 
has been observed at \mbox{$\beta=5.2$}~\cite{Farchioni:2004us}. 
The change between those two scenarios seems to happen
between these two values of~$\beta$. 

\begin{figure}[t]
  \centering
  \includegraphics[width=7.0cm]{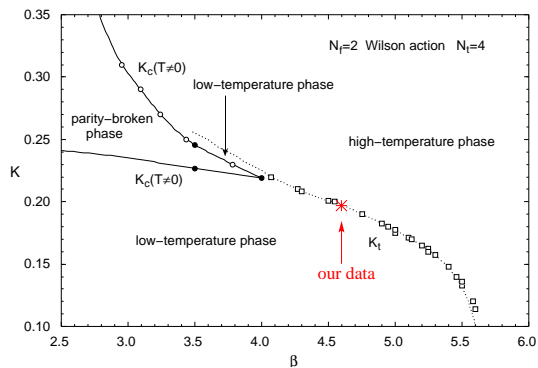}%
  \caption{The phase diagram with Wilson fermions at finite
    temperature as presented in \cite{Aoki:1995yf}. 
    We have added a point at $\beta=4.6$ where we have found 
    evidence for the Aoki phase.}
  \label{fig:fig4_Aoki}
\end{figure}

From the very beginning it was interesting to know where the Aoki
phase lies at finite temperature. Numerical
studies \cite{Aoki:1995yf,Aoki:1997fm} have given
some indications that the Aoki phase is completely embedded in the low
temperature phase as proposed by Aoki \etal \cite{Aoki:1995yf}. 
Their results for $N_f=2$ and $N_{\tau}=4$ 
are shown in \Fig{fig:fig4_Aoki}. However, the results left open the
question, whether the Aoki phase will join the 
thermal transition line $K_T(\beta,N_{\tau})$ as it extends
further in $\beta$ for a given thermal lattice extension~$N_{\tau}$.

To be specific, the following observations were made by Aoki \etal
\cite{Aoki:1995yf} 
for $N_f=2$ dynamical flavors: For $N_{\tau}=4$ at $\beta=3.5$ (on a
$8^3\times4$ lattice) a finite parity-flavor breaking condensate $\op$ was found
in a $\kappa$ interval where
$m_{\pi^{\pm}}=0$. Surprisingly, the cusp of the Aoki phase moved
slightly 
towards larger $\beta$ when $N_{\tau}$ is increased. In fact, going from
$N_{\tau}=4$ to $N_{\tau}=8$ resulted only in a shift from
$\beta=4.0$ to $\beta=4.2$.  At larger $\beta$ no evidence
for the Aoki phase --- at finite temperature --- has been found in unquenched
simulations so far.

In a finite-temperature study by the MILC collaboration on a
$12^3\times6$ lattice~\cite{Blum:1994eh} metastabilities of
plaquette values at ($\beta,\kappa$) =(4.8,0.19), (5.02,0.18) and
(5.22,0.17) were found. This observation 
could be compatible with the metastability at
(5.20, 0.1715) observed in~\cite{Farchioni:2004us}.

A possible scenario could be that the tip of the cusp simply stops
moving and turns into a 
first order line along which $m_{\pi} \ne 0$, separating phases with
$m_q<0$ (at high $\kappa$) from $m_q>0$ (at low $\kappa$)
\cite{Creutz:1996bg,Ukawa_at_Shuzenji}.  

What happens to the first order phase transition line at weaker coupling?
Is it universal and how does this affect the continuum limit? 
Does the transition extend to $\beta \to \infty$ as a single first
order line? Or does it split again giving room for a new
\emph{Aoki-like} phase enclosed? Can the Aoki phase extend
towards $\beta \to \infty$ being enclosed by two second order lines
using improved actions? 

\section{Numerical results}

\begin{figure}[tb]
\centering \includegraphics[width=7cm]{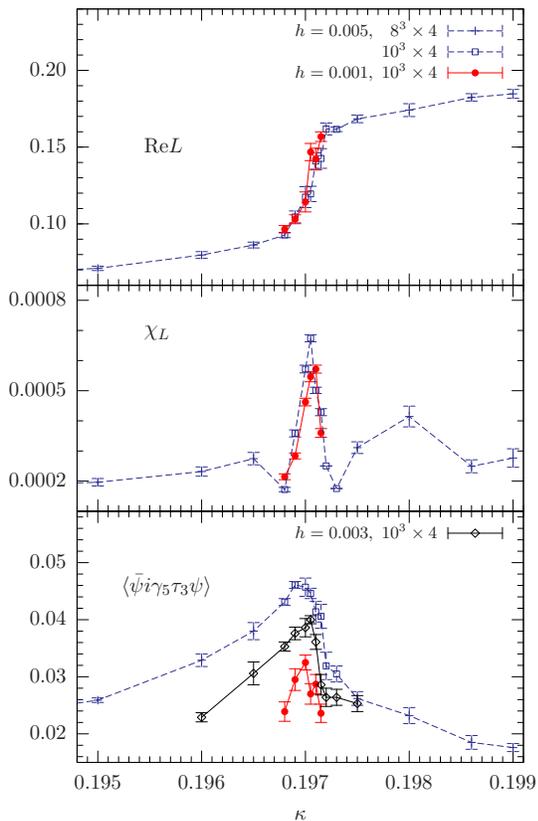}
\caption{The real part of the Polyakov loop, its susceptibility
  $\chi_L$ and the order parameter $\op$ as a function of $\kappa$ at
  \mbox{$h=0.005$} and 0.001 fixed. The lower panels shows $h=0.003$
  additionally. The data  have been obtained from simulations on a 
  $8^3\times4$ and a $10^3\times4$ lattice at $\beta=4.6$. Throughout the
  same symbols are used.}  
\label{fig:observables}
\end{figure}

\begin{figure}[tb]
\centering \includegraphics[width=7cm]{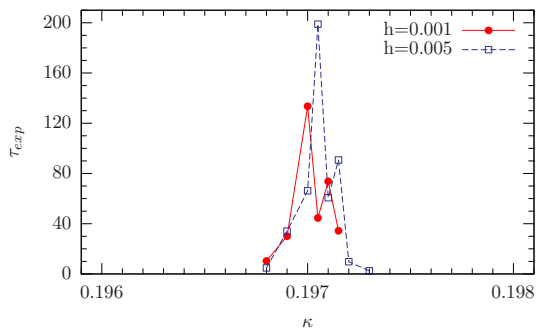}
\caption{The (exponential) autocorrelation time $\tau_{exp}$ 
  (in terms of HMC trajectories) of the real part of the Polyakov loop 
  is shown as a function of $\kappa$ for two different values of $h$. 
  The lattice size is $10^3\times4$.} 
\label{fig:autocorrelation}
\end{figure}

\begin{figure*}
  \newcommand{\len}{16cm} \centering
  \includegraphics[width=\len]{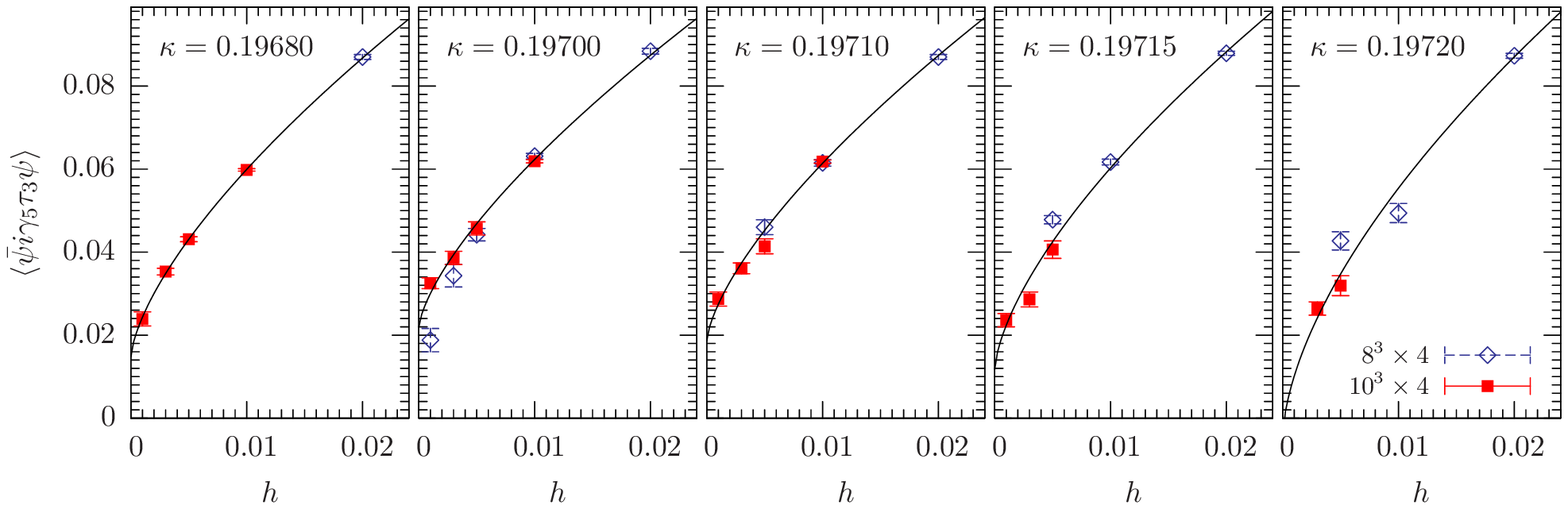}\\*[0.3cm]
  \includegraphics[width=\len]{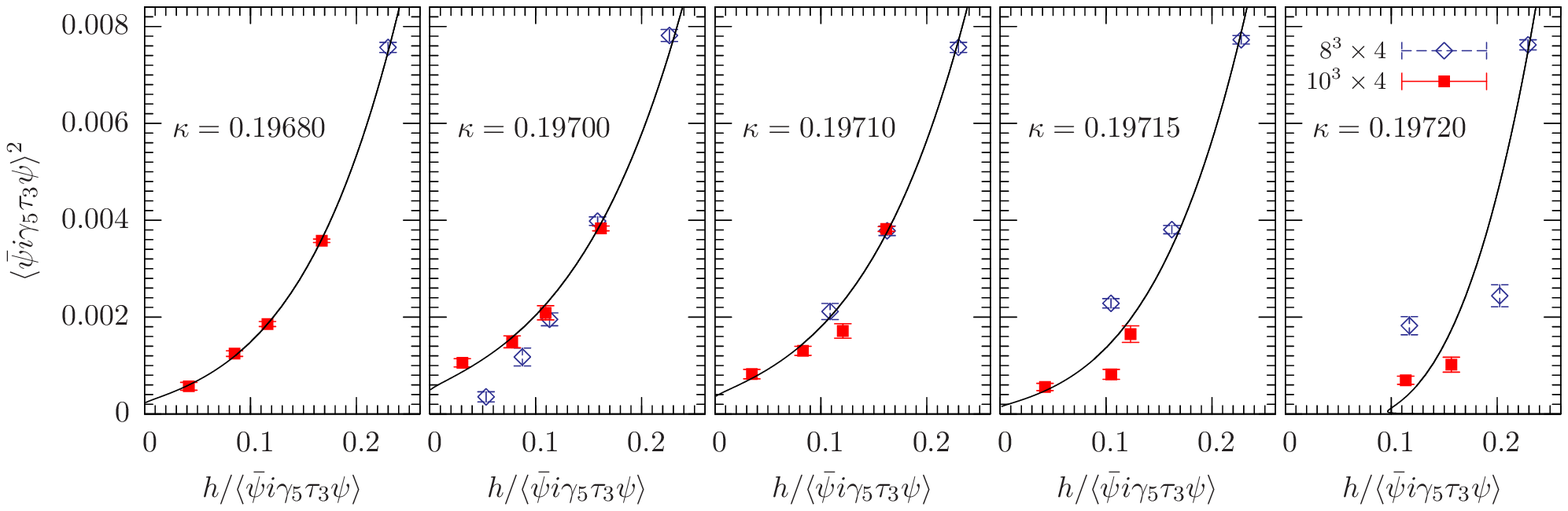}
  \caption{The upper panels show the order parameter $\op$ as a
  function of $h$ at five values of $\kappa$. The lines are fits to
    the data as described in the text. The lower panels
    show \emph{Fisher plots} of the order parameter  
    and the corresponding fitting function at the same values
    of $\kappa$ as above.  Throughout the same symbols are used.}
  \label{fig:op_h}
\end{figure*}

\begin{table*}
\renewcommand{\arraystretch}{0.6}       
\setlength{\tabcolsep}{4mm}             
\newcommand{\mc}{\multicolumn}
\centering{\footnotesize
  \begin{tabular}{crrrrrrrrrr}
    \hline\hline 
 $\kappa$ &  \mc{2}{c}{$h=0.001$} &  \mc{2}{c}{$h=0.003$} & \mc{2}{c}{$h=0.005$}
          &  \mc{2}{c}{$h=0.010$} &  \mc{2}{c}{$h=0.020$} \\ 
\hline
 0.19680 &  $10^3\times4$ & 1000 &  $10^3\times4$ &  1000 &  $10^3\times4$ & 1100 &  $8^3\times4$ & 1000 &  $8^3\times4$ &  90 \\ 
 0.19690 &  $10^3\times4$ & 1600 &  $10^3\times4$ &  2000 &  $10^3\times4$ & 1600 &  $8^3\times4$ & 1500 &  $8^3\times4$ &  60 \\ 
 0.19700 &  $10^3\times4$ &  2500 &  $10^3\times4$ & 4100 &  $8^3\times4$ & 3900 &  $10^3\times4$ & 1000 &  $8^3\times4$ &  150 \\ 
 0.19705 &  $10^3\times4$ & 1900 &  $10^3\times4$ & 1800 &  $10^3\times4$ & 4800 &  $10^3\times4$ & 1700 &  $8^3\times4$ &  400 \\ 
 0.19710 & $10^3\times4$  &  3200  & $10^3\times4$ & 3000 &  $10^3\times4$ & 3400 &  $10^3\times4$ & 1700 &  $8^3\times4$ &  150 \\ 
 0.19715 & $10^3\times4$  & 2000 & $10^3\times4$ & 1000 &
 $10^3\times4$ & 2500 &  $8^3\times4$ &  300 &  $8^3\times4$ &  200 \\
 0.19720 &   &  & $10^3\times4$ & 400 &  $10^3\times4$ & 400 &  $8^3\times4$ &  500 &  $8^3\times4$ &  300 \\ 
\hline\hline
\end{tabular}}
\caption{Statistics used for the final analysis at selected $\kappa$ at 
         $\beta=4.6$.}
\label{tab:stat}
\end{table*}
\begin{table*}
\renewcommand{\arraystretch}{0.3}         
\newcommand{\rb}[1]{\raisebox{1.5ex}[-1.5ex]{#1}}
\setlength{\tabcolsep}{1.8mm}             
\centering{\footnotesize
  \begin{tabular}{c@{\hspace*{0.6cm}}ccccccc@{\hspace*{1.0cm}}ccccccc}
    \hline\hline 
$\kappa$ & fit  &  A  &  B  & C  &  D  &  E & $\chi^2/\textsc{ndf}$ 
         & fit  &  A  &  B  & C  &  D  &  E & $\chi^2/\textsc{ndf}$ 
\\*[1ex] \hline
 &  \bf   1 &\bf 0.015(1) &\bf  1   &\bf 0.67(1) &\bf 0   &\bf 0 &\bf  0.24    
   &      2 &    0.016(1) &  1.07(7)&    0.69(2) &    0   &    0 &     0.24 \\
\rb{$0.19680$}
   &      3 &    0.026(1) &     1   &    0.68(1) & 0.1(1) &    0 &     0.23 
   &      4 &    0.015(1) &     1   &    0.68(1) &    0   & 2(2) &     0.19 \\*[0.25cm]

 & \bf   1 &\bf 0.021(1) &\bf  1   &\bf 0.70(1) &\bf 0   &\bf 0 &\bf  0.57 
   &      2 &    0.022(2) & 0.73(4) &    0.73(4) &    0   &    0 &     0.65 \\
\rb{$0.19690$} 
   &      3 &    0.022(2) &     1   &    0.72(2) & 0.2(2) &    0 &     0.64 
   &      4 &    0.022(1) &     1   &    0.71(1) &    0   & 4(4) &     0.59 \\*[0.25cm]
 &    \bf   1 &\bf 0.022(2) &\bf  1   &\bf 0.70(1) &\bf 0   &\bf 0 &\bf  2.48 
   &      2 &    0.026(1) & 1.5(2)  &    0.81(3) &    0   &    0 &     0.56 \\  
\rb{$0.19700$}
   &      3 &    0.026(2) &     1   &    0.76(3) & 0.6(2) &    0 &     0.59 
   &      4 &    0.025(2) &     1   &    0.72(1) &    0   & 11(4)&     0.87 \\*[0.25cm]
 &  \bf   1 &\bf 0.019(2) &\bf  1  &\bf 0.68(1) &\bf  0  &\bf 0 &\bf   4.0  
   &      2 &    0.023(4) & 1.2(3) &    0.74(7) &     0  &    0 &      4.3  \\
\rb{$0.19705$}
   &      3 &    0.022(3) &     1  &    0.72(4) & 0.3(3) &    0 &      4.3  
   &      4 &    0.022(3) &     1  &    0.70(2) &     0  &  5(6)&      4.2  \\*[0.25cm]
 &  \bf   1 &\bf 0.019(2) &\bf  1   &\bf 0.68(1) &\bf  0 &\bf 0 &\bf   2.48 
   &      2 &    0.017(2) &  0.9(3) &    0.66(9) &     0 &    0 &      3.54 \\ 
\rb{$0.19710$}
   &      3 &    0.017(5) &     1   &    0.67(4) &-0.1(4)&    0 &      3.49 
   &      4 &    0.017(4) &     1   &    0.67(2) &     0 & -4(8)&      3.31 \\*[0.25cm] 
 &    \bf   1 &\bf 0.012(3) &\bf  1   &\bf 0.66(1) &\bf  0 &\bf 0 &\bf   4.74 
   &      2 &    0.008(9) &  0.6(1) &    0.84(3) &     0 &    0 &      6.26 \\
\rb{$0.19715$}
   &      3 &    0.008(8) &     1   &    0.63(6) &-0.3(6)&    0 &      6.26 
   &      4 &    0.009(6) &     1   &    0.64(3) &     0 &-9(13)&      5.8  \\*[0.25cm] 
 &
    \bf   1 &\bf-0.005(6) &\bf  1   &\bf 0.61(2) &\bf  0 &\bf 0  &\bf 5.16  
   &      2 &    0.019(10)&    7(1) &     1.2(4) &     0 &    0  &    6.26  \\
\rb{$0.19720$}
   &      3 &    0.015(10)&     1   &     1(27)  &  3(41)&    0  &    0.65  
   &      4 &    0.014(2) &     1   &    0.77(3) &     0 & 60(8) &    0.14  \\
  \hline\hline 
  \end{tabular}}
  \caption{The parameters of the ansatz $~\sigma(h)=A+Bh^{C}+Dh+Eh^2~$ fitted
           to the data of $\op$ at $\beta=4.6$ without (fits labeled 1 and 2), 
           with linear,  or with quadratic corrections (labeled 3 and 4). 
           At each $h$ the result from the largest lattice was used in 
           the fit (for details see Table \ref{tab:stat}). Fixed parameters 
           are given by their value without indicating an error. In each case 
           the first fit (bold numbers) was used in \Fig{fig:op_h}.}
  \label{tab:fit}
\end{table*}

In this study we looked for the Aoki phase at finite temperature at
$\beta$ larger than in earlier studies.  
We performed numerical simulations 
at temporal lattice extension $N_{\tau}\equiv 4$  using dynamical unimproved
Wilson fermions. The Wilson fermion matrix $M_{W}$ was supplemented by
an explicit parity-flavor symmetry breaking term, \ie the two-flavor
fermion matrix was given by 
\begin{equation}
  \label{eq:fermionmatrix}
  M(h) = M_{W} + h\,i\gamma_5\tau^3.
\end{equation}
We simulated at $\beta=4.6$.
$\kappa$ was fine-tuned in the interval $[0.193,0.199]$ at 
fixed values of $h$. For each triple
$(\beta,\kappa,h)$ the order parameter 
$\op$, the real part of Polyakov loop $\ReL$ and its susceptibility
$\chiL$ were measured. These
observables are shown in \Fig{fig:observables}.  From the Figure
we see that $\ReL$ and $\chiL$ reveal a finite
temperature transition around \mbox{$\kappa=0.19705$}, while
the order parameter $\op$ signals an existence of the Aoki phase in
this $\kappa$ region. 

Note that we had found a vestigial region around $\kappa\approx
0.1984$ at $\beta=4.6$ in our previous zero-temperature 
study~\cite{Ilgenfritz:2003gw}.
Therefore, the Aoki phase seems to follow the finite-temperature
transition line and is shifted to somewhat lower $\kappa$ compared to the 
zero-temperature case. This shift, however, cannot be resolved in 
\Fig{fig:fig4_Aoki}. In any case, the Aoki phase extends inside a
longer cusp than seen in \cite{Aoki:1995yf}.

In the region of interest around \mbox{$\kappa=0.19705$} we
found rather long auto\-correlations in the Monte Carlo time histories
in contrast to other $\kappa$ values. The autocorrelation function has
been estimated for the Polyakov loop and we found large values for the
exponential $\tau_{exp}$ and the integrated autocorrelation time $\tau_{int}$
(in terms of HMC trajectories). Examples for $\tau_{exp}$ at
\mbox{$h=0.005$} and 0.001 are shown in \Fig{fig:autocorrelation}.

In order to study the Aoki phase, the limit  
\begin{equation}
\label{eq:lim_h_V}
  \op_{h=0}=\lim_{h\rightarrow 0}\lim_{V\rightarrow \infty}
  \langle\bar{\psi} i \gamma_{5} \tau^{3}\psi\rangle
\end{equation}
has to be taken.  Therefore, we made runs in the interval
$\kappa\in[0.1968,0.19720]$ at several values of $h\in[0.001,0.02]$ 
using spatial volumes $8^3$ and $10^3$.

In the upper panels of \Fig{fig:op_h} the results of those simulations
are shown together with fits to the data from the largest
lattice available at each $h$. For the fits we used the
ansatz~\cite{Ilgenfritz:2003gw}
\begin{equation}
  \sigma(h) = A + Bh^{C} + \ldots\quad,
  \label{eq:ansatz}
\end{equation}
where $\sigma\equiv\op$. The value of the order parameter at
\mbox{$h=0$} is given by the fit parameter $A$. The fits are
quite robust against the introduction of linear and quadratic
corrections. Details of the statistics achieved are presented in
\Tab{tab:stat}.  
We did not list the information refering to the smaller lattice size 
$8^3\times4$ in all those cases, where $10^3\times4$ data were
available. The results of all our fits are quoted in \Tab{tab:fit}.

The lower panels of \Fig{fig:op_h} show the same results as in the
upper panels, however, in a different parametrization (so called
\emph{Fisher plots}, see \eg \cite{Fisher,Hinnerk2,Ilgenfritz:2003gw}). This
parametrization has the advantage that the curves bend upwards, if
$\op$ is non-zero in the limit $h\rightarrow0$.
Hence, the finiteness of the intercept point can be read off better. 
In addition, it enables us to see directly if this 
intercept grows with the volume. 
At $\kappa=0.19700$ we have computed the order parameter at lower
values of $h$ also for the smaller lattice size $8^3\times4$. As expected,
a clear volume dependence is found, indicating that the existence of the 
Aoki phase becomes visible only for sufficient large volumes. 
The right most panels of \Fig{fig:op_h} show data at
($\kappa=0.19720$) to illustrated a case where the order parameter
$\op$ is found to vanish.

\section{Discussion} \label{sec:discussion}

By measuring the order parameter $\op$ and extrapolating it to $h=0$ we
conclude that there is a parity-flavor broken phase at finite temperature
($N_\tau=4$) at $\beta=4.6$ in the interval
$0.1968\le\kappa\le0.19715$.  In the zero-temperature case we  
had found \cite{Ilgenfritz:2003gw} that the Aoki phase seems to end in the
vicinity of $\beta=4.6$, $\kappa=0.1984$.  In other words, for a finite
temperature the endpoint of the Aoki phase is shifted towards larger $\beta$ 
and also to lower~$\kappa$.

This result differs from~\cite{Aoki:1995yf,Aoki:1997fm} where the
endpoint of the Aoki phase for $N_\tau=4$ was seen at $\beta=4.0$,
$\kappa=0.22$.  The finite-temperature transition line determined in
Refs.~\cite{Aoki:1995yf,Aoki:1997fm} crosses the interval $\beta=4.6$,
\mbox{$0.1968\le\kappa\le0.19715$} in which we see the Aoki
phase. Although we have studied the respective $\kappa$ interval in rather
small steps, we are not able to separate (the boundary of) the
Aoki phase from the thermal phase transition (cf.~\Fig{fig:fig4_Aoki}). 
We plan to extend our investigation to larger $N_{\tau}$ values in the
near future.

\section*{ACKNOWLEDGMENTS}
All simulations were performed on the Cray T3E at Konrad-Zuse-Zentrum f\"ur
Infor\-mations\-technik Berlin and on the IBM p690 system of
Norddeutscher Verbund f\"ur Hoch- und H\"ochstleistungsrechnen
(HLRN). A.~Sternbeck would like to thank the
DFG-funded graduate school GK~271 for financial support. E.-M.~I. is
supported by DFG through the Forschergruppe FOR 465 (Mu932/2-3).
M.~M.-P. acknowledges DFG support through SFB/TR~9.
E.-M.~I. wishs to send a special thank to S.~Aoki for the invitation to the
workshop \emph{Lattice QCD simulations via International Research Network},
Shuzenji, Japan, September 2004. The discussions there have
motivated the extension of our previous work to finite temperature.


\bibliographystyle{apsrev}

\end{document}